\documentclass[prb,twocolumn]{revtex4}
\usepackage{amsmath, amsgen,amstext,amsbsy,amsopn,amsthm, amssymb}

\newtheorem{thm}{Theorem}

\newtheorem{lem}{Lemma}

\theoremstyle{definition}

\newcommand{\half}{\mbox{$\frac{1}{2}$}}
\newcommand{\E}{\mathcal{E}}

\newcommand{\x}{{\bf r}}

\newcommand{\R}{\mathbb{R}}
\newcommand{\Z}{\mathbb{Z}}

\newcommand{\const}{{\rm const. \,}}

\newcommand{\K}{{\cal K}}
\newcommand{\X}{{\bf X}}
\newcommand{\Tr}{{\rm Trace}\, }
\newcommand{\vp}{\mathord{\hbox{\boldmath $\varphi$}}}

\begin{document}
\title{Superfluidity in Dilute Trapped Bose Gases}
\author{Elliott H.~Lieb}
\email{lieb@math.princeton.edu}
\affiliation{Department of Physics,
Jadwin Hall, Princeton University, 
P.~O.~Box 708, Princeton, New Jersey 08544}
\author{Robert Seiringer}
\altaffiliation{On leave from Institut f\"ur Theoretische Physik, 
Universit\"at 
Wien, Boltzmanngasse 5, A-1090 Vienna, Austria}
\email{rseiring@math.princeton.edu}
\affiliation{Department of Physics,
Jadwin Hall, Princeton University, 
P.~O.~Box 708, Princeton, New Jersey 08544}
\author{Jakob Yngvason}
\email{yngvason@thor.thp.univie.ac.at}
\affiliation{Institut f\"ur Theoretische Physik, 
Universit\"at 
Wien, Boltzmanngasse 5, A-1090 Vienna, Austria}

\date{\today}
\begin{abstract}
A commonly used theoretical definition of superfluidity in the ground
state of a Bose gas is based on the response of the system to an
imposed velocity field or, equivalently, to twisted boundary
conditions in a box.  We are able to carry out this program in the
case of a dilute interacting Bose gas in a trap, and we prove that a gas
with repulsive interactions is 100\% superfluid in the dilute limit in
which the Gross-Pitaevskii equation is exact.  This is the first
example in an experimentally realistic continuum model in which
superfluidity is rigorously verified.
\end{abstract}

\maketitle

\section{Introduction}

The phenomenological two-fluid model of superfluidity (see, e.g., \cite{TT}) 
is based on the
idea that the particle density $\rho$ is composed of two parts, the 
density $\rho_{\rm s}$ of the inviscid superfluid and the normal fluid density
$\rho_{\rm n}$.  If an external velocity field is imposed on the fluid
(for instance by moving the walls of the container) only the viscous normal
component responds to the velocity field, while the superfluid
component stays at rest.  In accord
with these ideas the superfluid density in the ground state is
often defined as follows \cite{HM}: Let $E_0$ denote the ground
state energy of the system in the rest frame and $E_0'$ the ground
state energy, measured in the
moving frame, when a velocity field ${\bf v}$ is imposed.  
Then for small ${\bf v}$
\begin{equation}\label{rhos}
    \frac{E_0'}N=\frac {E_0}N+({\rho_{\rm s}}/\rho)\half{ m} {\bf
    v}^2+O(|{\bf v}|^4) 
\end{equation} 
where $N$ is the particle number and $m$ the particle mass. At
positive temperatures the ground state energy should be replaced by
the free energy. (Remark: It is important here that (\ref{rhos}) holds
uniformly for all large $N$; i.e., that the error term $O(|{\bf
v}|^4)$ can be bounded independently of $N$.  For fixed $N$ and a
finite box, Eq.\ (\ref{rhos}) with $\rho_{\rm s}/\rho=1$ always holds
for a Bose gas with an arbitrary interaction if ${\bf v}$ is small
enough, owing to the discreteness of the energy spectrum
\cite{remark}.)  There are other definitions of the superfluid density
that may lead to slightly different results \cite{PS}, but this is the
one we shall use in this paper. We shall not dwell on this issue here,
since it is not clear that there is a \lq\lq one-size-fits-all\rq\rq\
definition of superfluidity. For instance, in the definition we use
here the ideal Bose gas is a perfect superfluid in its ground state,
whereas the definition of Landau in terms of a linear dispersion
relation of elementary excitations would indicate otherwise. We
emphasize that we are not advocating any particular approach to the
superfluidity question; our contribution here consists in taking one
standard definition and making its consequences explicit.

One of the unresolved issues in the theory of superfluidity is its
relation to Bose-Einstein condensation (BEC). It has been argued that
in general neither condition is necessary for the other (c.f., e.g.,
\cite{huang,giorgini,KT}). A simple example illustrating the fact 
that BEC is not necessary for superfluidity is the 1D hard-core Bose
gas. This system is well known to have a spectrum like that of an
ideal Fermi gas \cite{gira}, and it is easy to see that it is
superfluid in its ground state in the sense of (\ref{rhos}). On the
other hand, it has no BEC \cite{Lenard,pita}. The definition of the
superfluid velocity as the gradient of the phase of the condensate
wave function \cite{HM,baym} is clearly not applicable in such cases.

We do not give a historical overview of superfluidity because
excellent review articles are available \cite{leggett, baym}. While
the early investigations of superfluidity and Bose-Einstein
condensation were mostly concerned with liquid Helium 4, it has become
possible in recent years to study these phenomena in dilute trapped
gases of alkali atoms \cite{DGPS}. The experimental success in
realizing BEC in such gases has led to a large number of theoretical
papers on this subject. Most of these works take BEC for granted and
start off with the Gross-Pitaevskii (GP) equation to describe the
condensate wave function. A rigorous justification of
these assumptions is however a difficult task, and only very recently
BEC has been rigorously proved for a physically realistic many-body
Hamiltonian \cite{LS}. It is clearly of interest to show that 
superfluidity also holds in this model and this is what we accomplish 
here.  We prove that the ground state of a Bose
gas with short range, repulsive interaction is 100\% superfluid in the
dilute limit in which the Gross-Pitaevskii description of the gas is
exact. This is the limit in which the particle number tends to
infinity, but the ratio $Na/L$, where $a$ is the scattering length of
the interaction potential and $L$ the box size, is kept fixed. (The
significance of the parameter $Na/L$ is that it is the ratio of the
ground state energy per particle, $\sim Na/L^3$, to the lowest
excitation energy in the box, $\sim 1/L^2$.) In addition we show that
the gas remains 100\% Bose-Einstein condensed in this limit, also for
a finite velocity ${\bf v}$. Both results can be generalized from
periodic boxes to (non-constant) velocity fields in a cylindrical
geometry.

The results of this paper have been conjectured for many years, and it
is gratifying that they can be proved from first
principles.  They represent the first example of a rigorous
verification of superfluidity in an experimentally realistic
continuum model.

We wish to emphasize that in this GP limit the fact that there is
100\% condensation does not mean that no significant interactions
occur.  The kinetic and potential energies can differ markedly from
that obtained with a simple variational function that is an $N$-fold
product of one-body condensate wave functions. This assertion might
seem paradoxical, and the explanation is that near the GP limit the
region in which the wave function differs from the condensate function
has a tiny volume that goes to zero as $N\to \infty$. Nevertheless,
the interaction energy, which is proportional to $N$, resides in this
tiny volume.

\section{Setting and Main Results}

We consider a Bose gas with the Hamiltonian
\begin{equation}\label{ham}
H_{N}=-\mu\sum_{j=1}^N \nabla_j^2+
\sum_{1\leq i<j\leq N}v(\vert\x_{i}-\x_{j}\vert) \ ,
\end{equation}
where $\mu=\hbar^2/(2m)$ and the interaction potential $v$ is
nonnegative and of finite range. The two-body scattering length of $v$
is denoted by $a$. The Hamiltonian acts on totally symmetric functions
$\Psi$ of $N$ variables $\x_i=(x_i,y_i,z_i)\in
\K\subset \R^3$, where $\K$ denotes the cube $[0,L]^3$ of
side length $L$. (We could easily use a cuboid of sides
$L_1,L_2,L_3$ instead.) We assume periodic boundary 
conditions in all 
three coordinate directions.

Imposing an external velocity field ${\bf v}=(0,0,\pm|{\bf v}|)$ means
that the momentum operator ${\bf p}={-{\rm i}\hbar \nabla}$ is
replaced by by ${\bf p}-m{\bf v}$, retaining the periodic boundary
conditions. 
The Hamiltonian in the moving frame is thus 
\begin{multline}\label{hamprime}
H_N'  = -\mu\sum_{j=1}^N \big(\nabla_j + {\rm i} \vp/L\big)^2+
\sum_{1\leq i<j\leq N}v(\vert\x_{i}-\x_{j}\vert) \ ,
\end{multline}
where  $\vp=(0,0,\varphi)$ and
the dimensionless phase $\varphi$ is connected to 
the velocity ${\bf v}$ by
\begin{equation}
\label{varphi}\varphi=\frac{\pm|{\bf v}|Lm}\hbar\ . 
\end{equation}

Let $E_0(N,a,\varphi)$ denote the ground state energy of (\ref{hamprime})
with periodic boundary conditions. Obviously it is no
restriction to consider only the case $-\pi\leq \varphi\leq \pi$,
since $E_0$ is periodic in $\varphi$ with period $2\pi$.  For $\Psi_0$
the ground state of $H_N'$, let $\gamma_N$ be its one-particle reduced
density matrix
\begin{multline}
\gamma_N(\x,\x')= N \int_{\K^{N-1}} \Psi_0(\x,\x_2,\dots,\x_N) 
\\ \times \Psi_0^{\ast}(\x',\x_2,\dots,\x_N)\, d\x_2\cdots d\x_N \ .
\end{multline}
We are interested in the {\it Gross-Pitaevskii} (GP) limit
$N\to\infty$ with $Na/L$ fixed. We also fix the box size $L$. This
means that $a$ should vary like $1/N$ which can be achieved by writing
$v({\bf r})=a^{-2}v_1({\bf r}/a)$, where $v_1$ is a fixed potential
with scattering length 1, while $a$ changes with $N$.

\begin{thm}[Superfluidity of homogeneous gas]\label{T1}
For $|\varphi| \leq \pi$
\begin{equation}\label{i}
\lim_{N\to\infty} \frac{E_0(N,a,\varphi)}N = 4\pi\mu  a\rho + 
\mu\frac{\varphi^2}{L^2} 
\end{equation} 
in the limit $N\to \infty$ with $Na/L$ and $L$ fixed. Here $\rho=N/L^3$, 
so $a\rho$ is fixed too.
In the same 
limit, 
for $|\varphi| < \pi$, 
\begin{equation}\label{ii}
\lim_{N\to\infty} \frac 1N\, \gamma_N(\x,\x')=
\frac 1{L^3}
\end{equation}
in trace class norm, i.e., $\lim_{N\to\infty} \Tr \big[\,\big|\gamma_N/N - |L^{-3/2}\rangle\langle L^{-3/2}|\, \big|\,\big]=0$. 
\end{thm}

Note that, by the definition (\ref{rhos}) of $\rho_{s}$ and Eq.\
(\ref{varphi}), Eq.\ (\ref{i}) means that $\rho_{s}=\rho$, i.e., there
is 100\% superfluidity. For $\varphi=0$, Eq.\ (\ref{i}) was first
proved in \cite{LY1998}.  Eq.\ (\ref{ii}) for $\varphi=0$ is the BEC
proved in \cite{LS}.

{\it Remarks.}
1. By a unitary gauge transformation,
\begin{equation}\label{gauge}
\big(U\Psi\big)(\x_1,\dots,\x_N)= e^{{\rm i}\varphi(\sum_i z_i)/L}\, 
\Psi(\x_1,\dots,\x_N) \ , 
\end{equation}
the passage from (\ref{ham}) to (\ref{hamprime})  is equivalent to replacing periodic boundary
conditions in a box by the 
{\it twisted
boundary condition}
\begin{equation}\label{twist}
\Psi(\x_1 + (0,0,L), \x_2, \dots, \x_N)= e^{{\rm i}\varphi} \Psi(\x_1, 
\x_2, \dots, \x_N)
\end{equation}
in the direction of
the velocity field, while retaining the original Hamiltionian (\ref{ham}).

2. The criterion $|\varphi|\leq\pi$ means that $|{\bf v}|\leq \pi\hbar/(mL)$. The corresponding energy $\half m ( \pi\hbar/(mL) )^2$ is the gap in the excitation spectrum of the one-particle Hamiltonian in the finite-size system. 

3. The reason that we have to restrict ourselves to $|\varphi|<\pi$ in the
second part of Theorem~\ref{T1} is that for $|\varphi|=\pi$ there are
two ground states of the operator $(\nabla+{\rm i}\vp/L)^2$ with periodic
boundary conditions.  All we can say in this case is
that there is a subsequence of $\gamma_N$ that converges to a density
matrix of rank $\leq 2$, whose range is spanned by these two functions.

\bigskip
Theorem~\ref{T1} can be generalized in various ways to a physically
more realistic setting. As an example, let ${\cal C}$ be a finite cylinder
based on an annulus centered at the origin. Given a bounded, real
function $a(r,z)$ let $A$ be the vector field (in polar coordinates)
$A(r,\theta,z)=\varphi a(r,z) \widehat e_\theta$, where $\widehat
e_\theta $ is the unit vector in the $\theta$ direction. We also allow
for a bounded external potential $V(r,z)$ that does not depend on
$\theta$.

Using the methods of Appendix A in \cite{lsy1}, it is not difficult to
see that there exists a $\varphi_0>0$, depending only on ${\cal C}$ and
$a(r,z)$, such that for all $|\varphi|<\varphi_0$ there is a unique
minimizer $\phi^{\rm GP}$ of the Gross-Pitaevskii functional
\begin{multline}\label{defgp}
\E^{\rm 
GP}[\phi]=\int_{\K}\Big(\mu\big|\big(\nabla+{\rm i}A(\x)\big)
\phi(\x)\big|^2 \\ + V(\x) 
|\phi(\x)|^2 + 4\pi\mu N a 
|\phi(\x)|^4\Big)d^3\x 
\end{multline}
under the normalization condition $\int|\phi|^2=1$. This minimizer
does not depend on $\theta$, and can be chosen to be positive, for the
following reason: The relevant term in the kinetic energy is $T=
-r^{-2}[\partial/\partial \theta + {\rm i}\varphi\, r\, a(r,z)]^2$. If
$|\varphi\, r\, a(r,z)| < 1/2$, it is easy to see that $T\geq \varphi^2
a(r,z)^2$, in which case, without raising the energy, we can replace
$\phi$ by the square root of the $\theta$-average of $|\phi|^2$.  This
can only lower the kinetic energy \cite{anal} and, by convexity of
$x\to x^2$, this also lowers the $\phi^4$ term.

We denote the ground state energy of $\E^{\rm GP}$ by
$E^{\rm GP}$, depending on $Na$ and $\varphi$. 
The following Theorem \ref{T2} 
concerns the ground state energy $E_0$ of
\begin{multline}
H_N^{A}=\sum_{j=1}^N\Big[-\mu \big(\nabla_j+{\rm i}A(\x_j)\big)^2 +
V(\x_j)\Big]
\\ +\sum_{1\leq i<j\leq N}v(\vert\x_{i}-\x_{j}\vert) \ ,
\end{multline}
with Neumann boundary conditions on ${\cal C}$, and the one-particle reduced
density matrix $\gamma_N$ of the ground state, respectively. Different
boundary conditions can be treated in the same manner, if they are
also used in (\ref{defgp}).

{\it Remark.}  As a special case, consider a uniformly rotating
system. In this case $A(\x)=\varphi r \widehat e_\theta$, where $2
\varphi$ is the angular velocity. $H_N^A$ is the Hamiltonian in
the rotating frame, but with external potential $V(\x)+\mu A(\x)^2$
(see e.g. \cite[p.~131]{baym}).

\begin{thm}[Superfluidity in a cylinder]\label{T2} 
For $|\varphi|<\varphi_0$  
\begin{equation}\label{gpone}
\lim_{N\to\infty} \frac{E_0(N,a,\varphi)}N = E^{\rm GP}(Na,\varphi) 
\end{equation} 
in the limit $N\to \infty$ with $Na$ fixed. In the same limit, 
\begin{equation}
\lim_{N\to\infty} \frac 1N\, \gamma_N(\x,\x')=
\phi^{\rm GP}(\x)\phi^{\rm GP}(\x') 
\end{equation}
in trace class norm, i.e., $\lim_{N\to\infty} \Tr \big[\,\big|\gamma_N/N - |\phi^{\rm GP}\rangle\langle \phi^{\rm GP}|\, \big|\,\big]=0$. 
\end{thm}

In the case of a uniformly rotating system, where $2\varphi$ is the
angular velocity, the condition $|\varphi|< \varphi_0$ in
particular means that the angular velocity is smaller than the
critical velocity for creating vortices \cite{fetter}. 

{\it Remark.} 
In the special case of 
the curl-free vector potential $A(r,\theta)=\varphi r^{-1} \widehat
e_\theta$, i.e., $a(r,z)=r^{-1}$, one can say more about the role of $\varphi_0$.  In this case, there is a unique GP
minimizer for all $\varphi\not\in \Z+\half$, whereas there are two
minimizers for $\varphi\in \Z+\half$. Part two of Theorem \ref{T2}
holds in this special case for all $\varphi\not\in \Z+\half$, and
(\ref{gpone}) is true even for all $\varphi$.

\section{Proofs}

In the following, we will present only a proof of Theorem~\ref{T1} for
simplicity. Theorem~\ref{T2} can be proved using the same methods, and
additionally the methods of \cite{LS} to deal with the inhomogeneity
of the system.

Before giving the formal proofs, we outline the main ideas. The
strategy is related to the one in \cite{LS}, but requires substantial
generalizations of the techniques. A crucial element of the proof,
stated in Lemma \ref{L1} below, is the fact that the interaction
energy can be localized in small balls around each particle. This part
uses a Lemma of Dyson \cite{dyson}, and its generalization in
\cite{LY1998}, which converts a strong short range potential into a
soft potential. This Lemma can be also be applied to the case of an
external velocity field, i.e., a $U(1)$ gauge field in the kinetic term
of the Hamiltonian, owing to the \lq\lq diamagnetic inequaltity\rq\rq 
\cite{anal}. This inequaltity says that the additional gauge field increases the kinetic energy density. 

The second main part of the proof is the generalized Poincar\'e
inequality given in Lemma \ref{L2}. We recall that an essential
ingredient of the proof of Bose-Einstein condensation in \cite{LS} was
showing that the fact that the kinetic energy density is small in most
of the configuration space implies that the one-body reduced density
matrix is essentially constant. The difficulty comes from the fact
that the region in which the kinetic energy is small can, in
principle, be broken up into disjoint subregions, thereby permitting
different constants in different subregions. The fact that this does
not happen is the content of the generalized Poincar\'e inequality. In
the present case we have an additional complication coming from the
imposed gauge field. The old Poincar\'e inequality does not suffice;
one now has to measure the kinetic energy density relative to the
lowest energy of a free particle in the gauge field rather than to
zero. This is an essential complication. While the previous
(generalized) Poincar\'e inequality could, after some argumentation,
be related to the standard Poincar\'e inequality \cite{anal}, this new
one, with the gauge field, requires a totally new proof.

\begin{proof}[Proof of Theorem~\ref{T1}]

As in \cite{LY1998} we define $Y=(4\pi /3)\rho a^3$. Note that in the
limit considered, $Y\sim N^{-2}$. We first consider the upper
bound to $E_0$. Using the ground state $\Psi_0$ for $\varphi=0$ as a
trial function, we immediately get
\begin{equation}
E_0(N,a,\varphi)\leq \langle \Psi_0,  H_N' \Psi_0\rangle = 
E_0(N,a,0) + N\mu \frac{\varphi^2}{L^2} \ ,
\end{equation}
since $\langle\Psi_0, \nabla_i \Psi_0\rangle=0$. From \cite{lsy1} we 
know that $E_0(N,a,0)\leq 4\pi \mu N\rho a (1+\const Y ^{1/3})$, which has 
the right form as $N\to\infty$. 

For the lower bound to the ground state energy we need the following 
Lemma.

\begin{lem}[Localization of energy]\label{L1}
For all symmetric, normalized wave functions 
$\Psi(\x_{1},\dots,\x_{N})$ with periodic boundary conditions on 
$\K$, and for $N\geq Y^{-1/17}$, 
\begin{multline}\label{lowerbd}
\frac1N\langle\Psi, H_{N}'\Psi\rangle\geq \big(1-\const 
Y^{1/17}\big) \Big(4\pi\mu\rho 
a+ \\  
\mu \int _{\K^{N-1}}
d\X \int_{\Omega_{\X}}d\x_{1}\big|
\big(\nabla_1+{\rm i}\vp/L\big)\Psi(\x_{1},\X)\big\vert^2\Big) \ ,
\end{multline}   
where $\X=(\x_{2},\dots,\x_{N})$, $d\X=\prod_{j=2}^N d\x_j$, and 
\begin{equation}
\Omega_{\X}=\left\{\x_{1}:\ \min_{j\geq 2}\vert\x_{1}- 
\x_{j}\vert\geq R \right\} 
    \end{equation} 
with $R=a Y^{-5/17}$. 
\end{lem} 
 
\begin{proof} 
Since $\Psi$ is symmetric, the left side of 
(\ref{lowerbd}) can be written as
\begin{multline}    
\int_{\K^{N-1}} d\X \int_\K d\x_{1}
\Big[\mu\big|\big(\nabla_1+{\rm i}\vp/L\big)\Psi(\x_{1},\X)\big\vert^2
\\ +\half\sum_{j\geq 2}
v(\vert\x_{1}-\x_{j}\vert)|\Psi(\x_{1},\X)\vert^2\Big]\ .
\end{multline}
For any $\varepsilon>0$ and $R>0$ this is 
\begin{equation}
\geq \varepsilon T+(1-\varepsilon)(T^{\rm in}+I)+(1-\varepsilon)
T_{\varphi}^{\rm out} \ ,
\end{equation}
with
\begin{equation}\label{15}
T=\mu\int_{\K^{N-1}} d\X\int_\K 
d\x_{1}\big|\nabla_1\vert\Psi(\x_{1},\X)
\vert\big|^2 \ ,
\end{equation}
\begin{equation}\label{16}      
\quad T^{\rm in}=\mu
\int_{\K^{N-1}} d\X\int_{\Omega^c_{\X}} d\x_{1}\big|\nabla_1
\vert\Psi(\x_{1},\X)\vert\big|^2\ ,
\end{equation}  
\begin{equation}
T_{\varphi}^{\rm out}=\mu\int_{\K^{N-1}} d\X\int_{\Omega_{\X}} d\x_{1}
\big\vert\big(\nabla_1+{\rm i}\vp/L \big)
\Psi(\x_{1},\X)\big\vert^2\ ,
\end{equation}
and 
\begin{equation}
I=\half\int_{\K^{N-1}}
d\X\int_\K d\x_{1}\sum_{j\geq 2}
v(\vert\x_{1}-\x_{j}\vert)|\Psi(\x_{1},\X)\vert^2 \ .   
\end{equation}
Here
\begin{equation}\Omega^c_{\X}=\left\{\x_{1}:\ \vert\x_{1}- 
\x_{j}\vert<R\ \text{for some }j\geq 2\right\}
\end{equation}   
is the complement of $\Omega_{\X}$, and the diamagnetic inequality
$\vert(\nabla+{\rm i}\vp/L)f(\x)\vert^2\geq \left|\nabla|f(\x)|\right|^2$ 
has 
been used.  
The proof is completed by using the results of  \cite{LY1998} and 
\cite{LS} (see also \cite{LSSY}) which tell us that for
$\varepsilon=Y^{1/17}$ and $R=aY^{-5/17}$ 
\begin{equation}\label{abs}
\varepsilon T+(1-\varepsilon)(T^{\rm in}+I)\geq\big(1-\const 
Y^{1/17}\big)
4\pi\mu\rho a 
\end{equation}
as long as $N\geq Y^{-1/17}$. (This estimate is highly
non-trivial. Among several other things it uses a generalization of
Dyson's lemma \cite{dyson}.)
\end{proof}


The following Lemma \ref{L2} is needed for a lower bound on the second
term in (\ref{lowerbd}). It is stated for $\K$ the $L\times L\times
L$-cube with periodic boundary conditions, but it can be generalized
to arbitrary connected sets $\K$ that are sufficiently nice so that
the Rellich-Kondrashov Theorem (see \cite[Thm.~8.9]{anal}) holds on
$\K$. In particular, this is the case if $\K$ has the \lq cone
property\rq\ \cite{anal}. Another possible generalization is to
include general bounded vector fields replacing $\vp$, see
\cite{lsy02}.

If
$\Omega$ is any subset of $\K$ we shall denote
$\int_{\Omega}f^*(\x)g(\x)d\x$ by $\langle f,g\rangle_{\Omega}$ and
$\langle f,f\rangle_{\Omega}^{1/2}$ by $\Vert
f\Vert_{L^2(\Omega)}$. We also denote $\nabla+{\rm i}\vp$ by 
$\nabla_{\varphi}$ for
short.

\begin{lem}[Generalized Poincar\'e inequality]\label{L2}
For any $|\varphi|<\pi$ there  are constants 
$c>0$ and $C<\infty$  such that for all subsets 
$\Omega\subset\K$ and all functions $f$ on the torus $\K$  the 
following estimate holds:
\begin{multline}\label{poinc}
\Vert\nabla_\varphi
f\Vert_{L^2(\Omega)}^2\geq \frac{\varphi^2}{L^2}\|f\|_{L^2(\K)}^2 
+\frac c{L^2} \Vert 
f-L^{-3}\langle 1,f\rangle_{\K}
\Vert_{L^2(\K)}^2\\ -C\left(\|\nabla_\varphi f\|_{L^2(\K)}^2 + \frac 
1{L^2} 
\|f\|_{L^2(\K)}^2\right) \left(\frac{|\Omega|^c}{|\K|}\right)^{1/2} \ 
.
 \end{multline} Here $\vert\Omega^c\vert$ is the volume of
 $\Omega^c=\K\setminus\Omega$, the complement of $\Omega$ in $\K$.
\end{lem}

\begin{proof} 
We shall derive (\ref{poinc}) from a special form of this inequality
that holds for all functions that are orthogonal to the constant
function. Namely, for any positive $\alpha<2/3$ and some constants
$c>0$ and $\widetilde C<\infty$ (depending only on $\alpha$ and
$|\varphi|<\pi$) we claim that
\begin{multline}\label{poinc2}
\|\nabla_\varphi h\|_{L^2(\Omega)}^2 \\ \geq 
\frac{\varphi^2+c}{L^2}
\Vert h\Vert_{L^2(\K)}^2-\widetilde 
C\left(\frac{|\Omega^c|}{|\K|}\right)^\alpha\Vert\nabla_\varphi
h\Vert_{L^2(\K)}^2 \ ,
\end{multline}
provided 
$\langle 1,h\rangle_{\K} =0$. (Remark: Eq.~(\ref{poinc2}) 
holds also for $\alpha=2/3$, but the proof is slightly more 
complicated in that case. See \cite{lsy02}.) If (\ref{poinc2}) is 
known 
the 
derivation of (\ref{poinc}) is easy: For any $f$, the function
$h=f-L^{-3}\langle 1,f\rangle_{\K}$ is orthogonal to 
$1$. Moreover, 
\begin{eqnarray}\nonumber
&&\Vert\nabla_\varphi h\Vert^2_{L^2(\Omega)}\\ \nonumber &&=
\Vert\nabla_\varphi h\Vert^2_{L^2(\K)}-\Vert\nabla_\varphi 
h\Vert^2_{L^2(\Omega^c)}\nonumber\\   
&&=    
\Vert\nabla_\varphi f\Vert^2_{L^2(\Omega)}-
\frac{\varphi^2}{L^2}\vert\langle L^{-3/2},f\rangle_{\K}\vert^2 
\left(1+\frac{|\Omega^c|}{|\K|}\right)
\nonumber \\ &&\quad +2\frac{\varphi}{L}{\rm 
Re}\,\langle L^{-3/2},f\rangle_{\K}
\langle\nabla_\varphi f,L^{-3/2}
\rangle_{\Omega^c}\nonumber\\ \nonumber  
&&\leq \Vert\nabla_\varphi f\Vert^2_{L^2(\Omega)}-
\frac{\varphi^2}{L^2} \vert\langle L^{-3/2} ,f\rangle_{\K}\vert^2 
\\ \nonumber && \quad
+\frac{|\varphi|}{L}
\left(L \Vert\nabla_\varphi f\Vert_{L^2(\K)}^2+\frac 1L \Vert 
f\Vert_{L^2(\K)}^2\right) \left(\frac{|\Omega^c|}{|\K|}\right)^{1/2} \\ 
\label{quadrat1} 
\end{eqnarray}
and
\begin{multline}\label{quadrat2}
\frac{\varphi^2+c}{L^2}\Vert 
h\Vert_{L^2(\K)}^2=\frac{\varphi^2}{L^2}\left(  \Vert 
f\Vert_{L^2(\K)}^2- \vert\langle L^{-3/2} 
,f\rangle_{\K}\vert^2\right)\\ +
\frac c{L^2} \Vert f-L^{-3}\langle 1 ,f\rangle_{\K} 
\Vert_{L^2(\K)}^2 \ .
\end{multline}
Setting $\alpha=\half$, using $\|\nabla_\varphi h\|_{L^2(\K)}\leq 
\|\nabla_\varphi f\|_{L^2(\K)}$ in the last term in (\ref{poinc2}) 
and combining (\ref{poinc2}), (\ref{quadrat1}) and (\ref{quadrat2}) 
gives 
(\ref{poinc}) with $C=|\varphi|+\widetilde C$. 

We now turn to the proof of (\ref{poinc2}). For simplicity we set 
$L=1$. The general case  follows by scaling.  Assume that 
(\ref{poinc2}) 
is false. Then there exist sequences of constants $C_{n}\to\infty$,
functions $h_{n}$ with $\Vert h_{n}\Vert_{L^2(\K)}=1$ and 
$\langle 1,h_{n}\rangle_{\K} =0$, and domains $\Omega_{n}\subset\K$ such 
that
\begin{equation}\label{false}
\lim_{n\to\infty}\left\{
\Vert\nabla_\varphi
h_{n}\Vert_{L^2(\Omega_{n})}^2+C_{n}\vert\Omega_{n}^c\vert^\alpha\Vert\nabla_\varphi
h_{n}\Vert_{L^2(\K)}^2\right\}
\leq \varphi^2 \ .
\end{equation}
We shall show that this leads to a contradiction.

Since the sequence $h_{n}$ is bounded in $L^2(\K)$ it has a
subsequence, denoted again by $h_{n}$, that converges weakly to some
$h\in L^2(\K)$ (i.e., $\langle g,h_{n}\rangle_{\K}\to \langle
g,h\rangle_{\K}$ for all $g\in L^2(\K)$).  Moreover, by H\"older's
inequality the $L^p(\Omega_{n}^c)$ norm $\Vert \nabla_\varphi
h_{n}\Vert_{L^p(\Omega_{n}^c)}=(\int_{\Omega^c_{n}} \vert
h(\x)\vert^pd\x)^{1/p}$ is bounded by
$\vert\Omega_{n}^c\vert^{\alpha/2}\Vert\nabla_\varphi
h_{n}\Vert_{L^2(\K)}$ for $p=2/(\alpha+1)$.  From (\ref{false}) we
conclude that $\Vert \nabla_\varphi h_{n}\Vert_{L^p(\Omega_{n}^c)}$ is
bounded and also that $\Vert \nabla_\varphi
h_{n}\Vert_{L^p(\Omega_{n})}\leq\Vert \nabla_\varphi
h_{n}\Vert_{L^2(\Omega_{n})}$ is bounded. Altogether, $\nabla_\varphi
h_{n}$ is bounded in $L^p(\K)$, and by passing to a further
subsequence if necessary, we can therefore assume that $\nabla_\varphi
h_{n}$ converges weakly in $L^p(\K)$. The same applies to $\nabla
h_{n}$. Since $p=2/(\alpha+1)$ with $\alpha<2/3$ the hypotheses of the
Rellich-Kondrashov Theorem \cite[Thm~8.9]{anal} are fulfilled and
consequently $h_{n}$ converges {\it strongly} in $L^2(\K)$ to $h$
(i.e., $\Vert h-h_{n}\Vert_{L^2(\K)}\to 0$).  We shall now show that
\begin{equation}\label{lowersemi}
\liminf_{n\to\infty}\Vert\nabla_\varphi
h_{n}\Vert_{L^2(\Omega_{n})}^2\geq \Vert\nabla_\varphi
h\Vert_{L^2(\K)}^2 \ .
\end{equation}
This will complete the proof because the $h_{n}$ are normalized and
orthogonal to $1$ and the same holds for $h$ by strong
convergence. Hence the right side of (\ref{lowersemi}) is necessarily
$>\varphi^2$, since for $|\varphi|<\pi$ the lowest eigenvalue of
$-\nabla_\varphi^2$, with constant eigenfunction, is
non-degenerate. This contradicts (\ref{false}).

Eq.~(\ref{lowersemi}) is essentially a consequence of the weak lower
semicontinuity of the $L^2$ norm, but the dependence on $\Omega_{n}$
leads to a slight complication.  First, Eq.~(\ref{false}) and 
$C_{n}\to
\infty$ clearly imply that $\vert\Omega_{n}^c\vert\to 0$, because 
$\Vert\nabla_\varphi
h_{n}\Vert_{L^2(\K)}^2>\varphi^2$.  By choosing
a subsequence we may assume that
$\sum_{n}\vert\Omega_{n}^c\vert<\infty$.  For some fixed $N$ let
$\widetilde\Omega_{N}=\K\setminus\cup_{n\geq N}\Omega_{n}^c$. Then 
$\tilde\Omega_{N}\subset\Omega_{n}$ for $n\geq N$. 
Since $\Vert\nabla_\varphi
h_{n}\Vert_{L^2(\Omega_{n})}^2$ is bounded, $\nabla_\varphi
h_{n}$ is also bounded in $L^2(\widetilde\Omega_{N})$ and a 
subsequence 
of it converges weakly in $L^2(\widetilde\Omega_{N})$ to 
$\nabla_\varphi
h$. Hence 
\begin{multline}\label{lowersemi2}
\liminf_{n\to\infty}\Vert\nabla_\varphi
h_{n}\Vert_{L^2(\Omega_{n})}^2 \\ \geq 
\liminf_{n\to\infty}\Vert\nabla_\varphi
h_{n}\Vert_{L^2(\widetilde \Omega_{N})}^2\geq\Vert\nabla_\varphi
h\Vert_{L^2(\widetilde\Omega_{N})}^2  \ . 
\end{multline}
Since 
$\widetilde\Omega_{N}\subset \widetilde\Omega_{N+1}$ and 
$\cup_{N}\widetilde\Omega_{N}=\K$ (up to a set of measure zero), we 
can 
now let $N\to\infty$ on the right side of (\ref{lowersemi2}). By 
monotone convergence this converges to $\Vert\nabla_\varphi
h\Vert_{L^2(\K)}^2$. This proves (\ref{lowersemi}) which, as remarked 
above, contradicts
(\ref{false}). 
\end{proof}

We now are able to finish the proof of Theorem~\ref{T1}. From 
Lemmas~\ref{L1} and~\ref{L2} we infer that, for any symmetric $\Psi$ with 
$\langle \Psi,\Psi\rangle=1$ and for $N$ large enough,  
\begin{eqnarray} \nonumber 
&&\frac1N\langle\Psi,H_{N}'\Psi\rangle  \big(1-\const 
Y^{1/17}\big)^{-1}\\ \nonumber && \geq  4\pi\mu\rho a + \mu 
\frac{\varphi^2}{L^2} \\ \nonumber && 
\quad - C Y^{1/17} \Big(\frac 1{L^2} + \frac 1N 
\big\langle\Psi,\mbox{$ \sum_j$} (\nabla_j+{\rm i}\vp) \Psi\big\rangle\Big) 
 \\ \nonumber && \quad + \frac c{L^2}
\int_{\K^{N-1}}
d\X \int_{\K} d\x_{1}\Big|
\Psi(\x_1,\X)\\ &&\qquad\qquad -L^{-3}\big[\mbox{$ \int_\K$} d\x \Psi(\x,\X)\big] 
\Big|^2 \ ,\label{lowerbd2}
\end{eqnarray}   
where we used that $|\Omega^c|\leq \mbox{$\frac{4\pi}3$} N R^3= \const L^3 Y^{2/17}$. From this 
we can infer two things. First, since the kinetic energy, divided by 
$N$, is certainly bounded independent of $N$, as the upper bound 
shows, we get that
\begin{equation}
\liminf_{N\to\infty} \frac{E_0(N,a,\varphi)}N \geq 4\pi \mu \rho a + 
\mu \frac{\varphi^2}{L^2}
\end{equation}
for any $|\varphi|<\pi$. By continuity this holds also for
$|\varphi|=\pi$, proving (\ref{i}). (To be precise,
$E_0/N-\mu\varphi^2L^{-2}$ is concave in $\varphi$, and therefore stays
concave, and in particular continuous, in the limit $N\to\infty$.) 
Secondly, since the upper and the lower bound to $E_0$ agree in the
limit considered, the positive last term in (\ref{lowerbd2}) has to vanish in the limit. I.e., we get that for the ground state wave function
$\Psi_0$ of $ H_N'$
\begin{multline}
\lim_{N\to\infty} \int_{\K^{N-1}}
d\X \int_{\K} d\x_{1}\Big|
\Psi_0(\x_1,\X)\\ -L^{-3}\big[\mbox{$ \int_\K$} d\x \Psi_0(\x,\X)\big] 
\Big|^2 = 0 \ .
\end{multline}
This proves (\ref{ii}), since 
\begin{multline}
 \int_{\K^{N-1}}
d\X \int_{\K} d\x_{1}\Big|
\Psi_0(\x_1,\X)-L^{-3}\big[\mbox{$ \int_\K$} d\x \Psi_0(\x,\X)\big] 
\Big|^2 \\ = 1- \frac 1{NL^3} \int_{\K\times\K} \gamma(\x,\x') d\x d\x' 
\ ,
\end{multline}
and therefore $N^{-1}\langle L^{-3/2}|\gamma_N|L^{-3/2}\rangle \to 1$. As explained in \cite{LS,LSSY} this suffices for  the convergence $N^{-1}\gamma_N \to |L^{-3/2}\rangle \langle L^{-3/2}|$  in trace class norm. 
\end{proof}

\section{Conclusions} 
We have shown that a Bose gas 
with short range, repulsive interactions is both a 100\% superfluid and
also 100\% Bose-Einstein condensed in its ground state in the 
Gross-Pitaevskii limit where the parameter $Na/L$ is kept fixed as $N\to
\infty$. This is a simultaneous large $N$ and low density limit,
because the dimensionless density parameter $\rho a^3$ is here
proportional to $1/N^2$. If $\rho a^3$ is not zero, but small, a
depletion of the Bose-Einstein condensate of the order $(\rho
a^3)^{1/2}$ is expected (see, e.g., \cite{PeS}). Nevertheless,
complete superfluidity in the ground state, e.g. of Helium 4, is
experimentally observed. It is an interesting open problem to deduce
this property rigorously from first principles.  In the case of a 
one-dimensional 
 hard-core Bose gas superfluidity in the ground state is easy to show, but nevertheless there is no Bose-Einstein condensation at all, not even in the ground state \cite{Lenard,pita}.

\begin{acknowledgments}
E.H.L. was partially
supported by the U.S. National Science Foundation
grant PHY 98-20650. 
R.S. was supported by the Austrian Science Fund in the from of an Erwin Schr\"odinger fellowship.  
\end{acknowledgments}


\begin{thebibliography}{99}

\bibitem{TT} D.R.~Tilley and J.~Tilley,  {\it Superfluidity and 
Superconductivity}, third edition,
Adam Hilger, Bristol and New York (1990).

\bibitem{HM} P.C.~Hohenberg and P.C.~Martin, Ann.\ Phys.\ (NY) {\bf 34},
291 (1965).

\bibitem{remark}
The ground state with ${\bf v}=0$ remains an eigenstate of the
Hamiltonian with arbitrary ${\bf v}$ since its total momentum is zero. Its
energy is $\half m  N{\bf v}^2$ above the ground state energy for ${\bf
v}=0$. Since in a finite box the spectrum of the Hamiltonian for arbitrary
${\bf v}$ is discrete and the energy gap above the ground state is bounded
away from zero for ${\bf v}$ small, the ground state for ${\bf v}=0$ is at
the same time the ground state of the Hamiltonian with ${\bf v}$ if $\half
m N{\bf v}^2$ is smaller than the gap.

\bibitem{PS} N.V.~Prokof'ev and B.V.~Svistunov, 
Phys.\ Rev.\ B {\bf 61}, 11282 (2000).

\bibitem{huang} K. Huang, in: {\it Bose-Einstein Condensation}, A. Griffin, D.W. Stroke, S. Stringari, eds., Cambridge University Press, 31--50 (1995).

\bibitem{giorgini} G.E. Astrakharchik, J. Boronat, J. Casulleras, and S. Giorgini, arXiv:cond-mat/0111165 (2001).

\bibitem{KT} M.~Kobayashi and M.~Tsubota, arXiv:cond-mat/0202364 (2002).

\bibitem{gira} M. Girardeau, J. Math. Phys. {\bf 1}, 516 (1960). 

\bibitem{Lenard} A.~Lenard,
J. Math. Phys. {\bf 5}, 930 (1964). 

\bibitem{pita} L.\ Pitaevskii and S.\ Stringari, J. Low Temp. Phys. {\bf 85}, 377 (1991).

\bibitem{baym} G. Baym, in: {\it Math. Methods in Solid State and Superfluid Theory}, Scottish Univ. Summer School of Physics, Oliver and Boyd, Edinburgh (1969). 

\bibitem{leggett} A.J. Leggett, Rev. Mod. Phys. {\bf 71}, S318 (1999).



\bibitem{DGPS} F. Dalfovo, S. Giorgini, L. Pitaevskii, and S. Stringari, Rev. Mod. Phys. {\bf 71}, 463 (1999). 




\bibitem{LS} E.H.~Lieb and R.~Seiringer, Phys.\ Rev.\ Lett. {\bf 88},
170409 (2002).


\bibitem{LY1998}
E.H.~Lieb and J.~Yngvason, Phys.\ Rev.\ Lett. \textbf{80}, 2504 (1998).


\bibitem{lsy1}
E.H.~Lieb, R.~Seiringer, and J.~Yngvason,
Phys.\ Rev.\ A {\bf 61}, 043602 (2000).

\bibitem{anal} E.H.~Lieb and M.~Loss,  {\it Analysis}, second edition,
American Mathematical Society (2001).



\bibitem{fetter} A.L. Fetter and A.A. Svidzinsky, J. Phys.: Condens. Matter {\bf 13}, R135 (2001).


\bibitem{dyson} F.J. Dyson, Phys. Rev. {\bf 106}, 317 (1957).



\bibitem{LSSY}
E.H.~Lieb, R.~Seiringer, J.P.~Solovej, and J.~Yngvason, to appear in `Contemporary Developments
in Mathematics 2001', International Press. arXiv:math-ph/0204027 (2002).

\bibitem{lsy02}
E.H.\ Lieb, R.\ Seiringer, and J.\ Yngvason, arXiv: math.FA/0205088 (2002).

\bibitem{PeS}
C.J.~Pethick and H.~Smith, {\it Bose-Einstein Condensation in Dilute Gases},  
Cambridge University Press (2002).
\end{thebibliography}
\end{document}